# An Adaptive Similarity Measure to Tune Trust Influence in Memory-Based Collaborative Filtering

Mohammad Reza Zarei, Mohammad Reza Moosavi
Dep. of Computer Science and
Engineering, School of Electrical and Computer Engineering,
Shiraz University,
Shiraz, Iran
Email: mr.zarei@cse.shirazu.ac.ir, smmosavi@shirazu.ac.ir

*Abstract*— **The aim of the recommender systems is to provide relevant and potentially interesting information to each user. This is fulfilled by utilizing the already recorded tendencies of similar users or detecting items similar to interested items of the user. Challenges such as data sparsity and cold start problem are addressed in recent studies. Utilizing social information not only enhances the prediction accuracy but also tackles the data sparseness challenges. In this paper, we investigate the impact of using direct and indirect trust information in a memory-based collaborative filtering recommender system. An adaptive similarity measure is proposed and the contribution of social information is tuned using two learning schemes, greedy and gradient-based optimization. The results of the proposed method are compared with state-of-the-art memory-based and model-based CF approaches on two real-world datasets, Epinions and FilmTrust. The experiments show that our method is quite effective in designing an accurate and comprehensive recommender system.**

## 1. Introduction

With the ever-increasing popularity of social media, people have become more and more interested to use social networks in recent years. The contacts and communication between users construct a large valuable network that is presented as rich social information in different forms, including friendships (in Facebook and Doubans), trust (in Epinions), and followers (in Twitter) [1]. These social relationships establish the possibility of spreading ideas and opinions between connected users which is referred to as social contagion. In real life, friendship in social networks yields social influence which tends to have strong effects on changing personal manners [2]. Also, according to the findings of psychologists, humans tend to associate with their socially connected users and get influenced by them, known as homophily [3]. Therefore, analyzing the social information could reveal human preferences and behaviors.

Information obtained from social networks has widely motivated researchers in different areas such as recommender systems [4]. Recommender systems aid users to choose from various alternatives (or even provide new choices) by suggesting the most relevant items [5].

The most widely used approach for recommendation systems is collaborative filtering (CF) [6]. It tries to predict the preference of a user to a new item based on previously rated items. This approach employs the available user-item ratings to predict the rate of new items for each user. This technique is utilized by many websites and applications due to its effectiveness and simplicity [7]. On the other hand, they may face the challenge of data sparsity and cold start problems [8]. To overcome these challenges, social ties have been widely used in past few years, not only to prevent cold start and sparsity problems from deteriorating the performance of recommender systems [9] but also to increase the overall performance.

CF-based algorithms can be divided into two main approaches, Model-based CF and Memory-based CF. Model-based approaches try to learn a model from user-item ratings which will be used for rating prediction [10]. One of the most popular model-based techniques is low-rank Matrix Factorization (MF) [11]–[15]. The idea of MF methods is to represent users and their preferences with corresponding low-rank latent vectors and then train a prediction model by optimizing some objective functions over ratings [7]. Constraining the entries of latent vectors to be non-negative, Non-negative Matrix Factorization (NMF) [16] gives more meaningful decomposition with non-unique factors in comparison with other MF techniques [17]. Several methods have been proposed to improve the performance of traditional MF algorithms by using probabilistic terms. One of these methods is termed as Probabilistic Matrix Factorization (PMF) [13] proposed by Salakhutdinov and Mnih which attempts to perform well on sparse and large-scale datasets.

All of these MF approaches are to alleviate the need for accessing the whole data for each prediction after building the model. They are also capable of predicting the rate of all items and therefore provide the full coverage. On the other hand, they own poor performance in very sparse networks while their learning phase makes them complex to be implemented and also incapable of providing an immediate response to a new user after receiving her feedback [18].

Recently, with the prospering performance of neural network models in the fields of Computer Vision and Natural Language Processing, researchers have tried to utilize these models in the field of recommender systems. For instance, AutoRec [19] proposed by Sedhain et. al is an autoencoder framework for collaborative filtering. The severity of data sparsity is a challenging issue for training these models.

Memory-based CF approaches [20]–[23] are based on calculating the similarity between the users and/or between the items. To make rating prediction for query user, user-based collaborative-filtering methods find similar users and use their opinions in the prediction process [24]. Memory-based CF algorithms are simple and intuitive as well as possessing admissible accuracy. In spite of this, computing the similarity measure for users with few rated items in their profile is unreliable if not impossible. Since this problem is common in highly sparse data, Memory-based CF algorithms cannot response to every prediction request.

The poor performance of both model-based and memory-based CFs in the cold start and sparse data, motivated researchers to utilize various types of information for answering rate-prediction queries. Generally, the Information Retrieval (IR) systems use some kinds of techniques to extract hidden concepts for query refining [25]. These techniques known as query expansion utilize different sources of user data to enrich the query and enhance the rate-prediction. Specifically, in recommender systems, researchers attempt to use social information for improving the performance of model-based and memory-based CF approaches. Several social MF-based methods have been proposed to improve recommendation accuracy [26]–[28]. In particular, Ma et al. [29] proposed a probabilistic factor analysis framework (RSTE) which derives the final decision of the user from her own taste and her friends' favors by an ensemble parameter. In [27], a social regularized recommender system (SoReg) is introduced that presents a matrix factorization objective function with social regularization. Guo et al. [30] proposed a trust-based matrix factorization technique (TrustSVD) which builds on top of SVD++ [31] and factorizes both user-item matrix and user-user matrix in a regularized manner. In [7], Yang et al. proposed TrustMF, a personalized social method that fuses both rating data and trust data. This method trains two models independently (Truster and Trustee) and combines them to generate the approximation of real ratings.

Social memory-based CFs [32]–[35] usually utilize social information by considering both the rates of trusted users and the rates of similar users to tackle data sparsity and cold start problem and also improve the performance of predictions. Massa and Avesani proposed an approach that uses a local trust metric called Moletrust to propagate and estimate trust [34]. In this method named MT, they used this trust measure instead of similarity value to response to prediction requests. In [36], Chowdhurry et al. proposed Trust-based Enhanced Collaborative Filtering (TCF) that uses the rates of trusted neighbors if the rates of similar users are not available. Guo et al. proposed a trust-based approach called "Merge" [37]. This method merges the co-rated items of trusted users and the merged rates are used to find the similarity between the users. The impact of directly trusted users are investigated in [35], but the method does not provide trust propagation and learning scheme for tuning the similarity measure.

To put in a nutshell, trust information could be integrated into recommender systems to decrease the impact of data sparsity on recommendation performance. In comparison with model-based CFs, memory-based algorithms have better performance in highly sparse data whereas their less coverage makes them impotent in predicting every rating. Trust information is useful for social memory-based approaches to increase prediction coverage.

Since the number of people trusted by a user is limited, more levels of trust (i.e., the information of indirectly trusted users) may be exploited to enhance the prediction performance and expand the rate-prediction query. Utilizing trust propagation, the amount of informative data from trusted users would be increased, impacting both prediction coverage and accuracy. It should be mentioned that the impact of using indirect trusts is not well studied in the literature to our knowledge.

In this paper, we introduce a memory-based collaborative-filtering method for rate-prediction. A similarity measure is proposed to expand the query user rates by integrating social ties. We also investigate the effect of trust propagation on social ties into our method. The proposed algorithm is tuned to optimize the usage of social relations as well as examining the effectiveness of trusted users' opinion regarding the prediction accuracy and coverage. We propagate the trust and learn the level of its influence to gain the maximum benefit from the viewpoint of indirect trusted users. At the end, we show that our method performs well in comparison with the state-of-the-art model-based and memory-based CF approaches.

This work makes the following contributions:
- Proposing a memory-based Collaborative Filtering algorithm.
- Proposing an adaptive similarity measure that determines the similarity of users incorporating information of social ties.
- Learning the coefficient parameters of the similarity measure to tune the influence of trust propagation.
- Analyzing the impact of social ties in rate-prediction based on accuracy and coverage.

## 2. METHODOLOGY

The proposed CF-based algorithm predicts the preference of a query user for a target item. The query user denotes who we want to predict for and the scorers are all of those users that have rated for the target item. Here, we first define the notations and rate-prediction based on the similarity between query user and other users. Then, the fusion of informative social ties (direct and indirect trusted users) is discussed. The prediction process utilizes the rate values of the users who have already rated the target item.

### 2.1. DEFINITIONS AND OVERVIEW

Assume that for a set of users $U = \{u_i | i = 1, ..., m\}$ and item set $I = \{i_j | j = 1, ..., n\}$, a set of ratings $R = \{r_{ij} | 1 \leq i \leq m \wedge 1 \leq j \leq n \wedge r_{min} \leq r_{ij} \leq r_{max}\}$ is given, where $r_{ij}$ denotes the rate that user $i$ gives to item $j$, $r_{min}$ and $r_{max}$ specify the minimum and maximum rate values that can be assigned to an item representing the lowest and highest interest, respectively. The prediction of a user's rate (query or active user) for a not yet rated item (target item) is done based on the existing rates of other users (scorers). User-based collaborative-filtering recommender systems are constructed by means of a similarity metric that quantifies the similarity of users based on their rates.

The similarity of users based on a single item, $i$, could be simply defined according to the closeness of their rate values. Suppose that two users $q$ and $s$, both rated the item $i$ (i.e, $r_{si} \in R \wedge r_{qi} \in R$). The closeness of these rates $RS(r_{qi}, r_{si})$ is calculated as:

$$RS(r_{qi}, r_{si}) = 1 - (|r_{qi} - r_{si}|/(r_{max} - r_{min})) \quad (1)$$

where the normalization factor $r_{max}$-$r_{min}$ is the maximum possible difference in rate values. For example, in a dataset that the items are rated using integer values of the range [1, 5], the maximum possible difference between the two rates is 4.

The similarity between user $q$ and user $s$ over all rated items is calculated as:

$$sim(q, s) = \frac{\sum_{i \in (I_q \cap I_s)} RS(r_{qi}, r_{si})}{|I_q \cap I_s|} \quad (2)$$

where $I_q = \{i | r_{qi} \in R\}$ and $I_s$ are the lists of items rated by user $q$ and user $s$, respectively. This similarity measure calculates the average closeness of rate values for all co-rated items.

The predicted rate of item *t* (target item) for user *q* (query user) is computed as the weighted average of previous rates of the item *t* by scorers. The weight factor is the similarity between the query user and each scorer. The predicted rate is calculated as:

$$\hat{r}_{qt} = \frac{\sum_{s \in U_t}[sim(q,s) \times r_{st}]}{\sum_{s \in U_t} sim(q,s)} \quad (3)$$

where $U_t = \{j \mid r_{jt} \in R\}$ is the list of scorers.

Obviously, in the real-world, user rating data is severely sparse, which means that a typical user does not declare her idea about most of the items. In other words, converting the set *R* to a matrix result in $r_{ij} = $ *nan* for many i and j values. This issue challenges the appropriateness of simple similarity measures such as equation (3). This drawback can be alleviated by the adaptation of an appropriate similarity measure.

## 2.2. SOCIAL RECOMMENDER SYSTEM

In equation (3), $\hat{r}_{qt}$ is simply calculated based on the similarity of rates of each scorer and the query user. Because of the data sparseness, few rates are available for the query user, which means that comparing the rates of query user and a scorer does not provide adequate information. The information from query user can be expanded by defining a set of user groups (or clusters) that are similar to the query user. These subsets of the users can be specified using various information such as homophily (race, gender, etc. based on profiles of users), the geographical proximity of users, social information, etc.

Here in this paper, we use social ties to enrich the available data from the query user. To do this, equation (2) is reformulated as a similarity measure between the active user and a set of users. The average similarity of a user *u* and a set of users *V* is defined as:

$$Sim(u,V) = \frac{\sum_{x \in V(I_x \cap I_u \neq \emptyset)} sim(u,x)}{|\{x \in V(I_x \cap I_u \neq \emptyset)\}|} \quad (4)$$

where $V \subseteq U$. The denominator of the equation (4) indicates the number of users, each of which has at least one co-rated item in common with the user *u*.

Generally, for a query user *q*, we can determine a set of user groups or clusters $V_1, ..., V_{g-1}$. Now the similarity of each scorer can be calculated not only with the query user but also with these groups. The query user itself can be considered as a special cluster consisting of one user $V_0=\{u_q\}$. A weighted average of similarities is used to predict the rate of item *t* by the user *q*:

$$\hat{r}_{qt} = \frac{\sum_{s \in U_t}(\sum_{i=0}^{g-1} \omega_i \times Sim(s,V_i)) \times r_{st}}{\sum_{s \in U_t} \sum_{i=0}^{g-1} \omega_i \times Sim(s,V_i)} \quad (5)$$

where $\omega_i$ is the weight of the group *i*. The appropriate weights could be the adaptation by means of learning techniques which is discussed in Subsection 2.4.

Using social ties, for each user *q*, the set $T_q$ is defined consisting of the users that are labeled as trusted by the user *q*. To incorporate the social relations in the prediction process, we calculate the similarity between the set of users trusted by user *q* and the scorer *s* as $Sim(s,T_q)$ using equation (4).

To integrate the resulted similarity from social relations into the rate-prediction process, equation (5) is simplified to:

$$\hat{r}_{qt} = \frac{\sum_{s \in U_t}[[\omega_0 sim(q,s) + \omega_1 Sim(s,T_q)] \times r_{st}]}{\sum_{s \in U_t}[\omega_0 sim(q,s) + \omega_1 Sim(T_q,s)]} \quad (6)$$

where $\omega_1 = (1 - \omega_0) \in [0,1]$. Clearly, this equation fuses the similarity of trusted users and the similarity of query user with each scorer. The contribution of similarity with trustees is controlled by the weight parameter $\omega_1$, which is adapted in the experiments as described in Section 3.

## 2.3. TRUST PROPAGATION

The query could be further expanded by means of trust propagation to improve rate-prediction accuracy. To exploit the impact of using trust propagation in the improvement of rate-prediction, we use the proposed similarity measure in equation (4) to calculate the similarity between indirectly trusted users and scorer *s*. Considering the trust network as a directed graph, for a query user *q*, an indirect trusted user is a user that has the distance of length 2 with *q*. Hence, for the user q, the set of indirectly trusted users is defined as:

$$T'_q = \{x' | x' \in T_x \cdot \forall x \in T_q\} \tag{7}$$

The similarity of indirect trusted users with each scorer, $Sim(s, T'_q)$, is applied to equation (6) with a coefficient parameter $\omega_2$ which controls its contribution, similar to the previously defined coefficients. As a result, the prediction formula (6) is rewritten as:

$$\hat{r}_{qt} = \frac{\sum_{s \in U_t}[\omega_0 sim(q,s) + \omega_1 Sim(s,T_q) + \omega_2 Sim(s,T'_q)] \times r_{st}}{\sum_{s \in U_t} \omega_0 sim(q,s) + \omega_1 Sim(s,T_q) + \omega_2 Sim(s,T'_q)} \tag{8}$$

where $\Omega = \{\omega_0, \omega_1, \omega_2\}$ are the coefficient parameters of the range $[l, u]$ and should be tuned as explained in the next subsection. The parameter tuning adjusts the relative influence of query user ($sim(q,s)$), directly trusted users ($Sim(s,T_q)$) and indirectly trusted users ($Sim(s,T'_q)$).

### 2.4. THE LEARNING SCHEME

In this subsection, our aim is to introduce two learning schemes in an attempt to maximize the prediction performance by optimizing the contribution parameters $\Omega$. The first one is a simple greedy algorithm and the second one is an optimization procedure inspired by the hill-climbing search method.

In the greedy scheme, neighbor solutions are defined as two sets of parameters' values that vary in the value of just one parameter. The basic component of the learning scheme is an algorithm that provides the answer to the following question:

*What is the best value of a parameter assuming that the value of all other parameters is given and fixed?*

---

1: **function** GREEDY_PARAM_TUNING ($\Omega$, *i*, *l*, *u*, $R_T$, $R_V$, *T*) **returns** *ω*
   ▷ **Input:** $\Omega$, set of the parameters and their values; $\omega_i$ is the parameter to be tuned, and other elements of $\Omega$ are fixed and invariant during the procedure; *l*, *u*: lower and upper bound for the parameter $\omega_i$, $R_T$: train rates, $R_V$: validation rates, *T*: trust network
   ▷ **Output:** *ω*, the tuned value for parameter $\omega_i$
2: **for** tuning_iteration *j* = 1 to *n* **do**
3:     *param_list* = SPLIT_RANGE(*l*, *u*)     ▷ produce the list of candidate values to be tested for parameter by splitting the given range of the parameter into ten equal parts
4:     *min_mae* = +∞     ▷ initialization
5:     for each parameter *p* in *param_list*
6:         *mae* = CALCULATE_MAE_ON_VALIDATION_DATA (*p*, $R_T$, $R_V$, *T*)
7:         **if** *mae* < *min_mae* **then**
8:             *min_mae* = *mae*
9:             *best_param* = *p*
10:     *l* = param_list[index(*best_param*)-1]     ▷ new lower bound for *param_list*
11:     *u* = param_list[index(*best_param*)+1]     ▷ new upper bound for *param_list*
12: **return** *best_param*

---

**Figure 1: Parameter tuning algorithm using the greedy scheme**

The algorithm starts with an initial solution to the problem (for example, $\omega_i = 0.5 \cdot \forall i$) and attempts to improve the solution by adjusting the weight of one parameter in each step. In each step, one of the parameters is selected and tuned. The value range of the parameter is split into ten equal sub-ranges and the effectiveness of each boundary value is measured by means of the prediction performance metric. Based on this feedback, the value range of the parameters is narrowed down. The algorithm of tuning the value of a parameter is presented in Figure 1. This algorithm is used to adjust the value of all parameters in $\Omega$. The algorithm uses two subsets of the rates as training set, $R_T$, and validation set $R_V$.

It must be noted that the learning procedure of each parameter is repeated to tune the parameters with respect to the new values of the other parameters. That is why the second pass and subsequent passes over the parameters can improve the performance. The value of a parameter is the best in the sense that it results in the best prediction performance. This way, the overall learning algorithm consists of several iterations, each of which deals with tuning the value of all parameters one after the other. In experiments, as a mechanism to prevent overfitting, we stop after a pre-specified number of iterations. The experiments of tuning and utilizing these parameters are explained in Section 3.

The second scheme is an iterative gradient-based optimization procedure. We use a hill-climbing approach to tune parameters in $\Omega$ in order to minimize the following objective function:

$$\mathcal{F} = \frac{1}{2|R_v|} \sum_{(i,j) \in R} (r_{ij} - \hat{r}_{ij})^2 \tag{9}$$

where $R_v = \{(i,j) | r_{ij} \neq \emptyset\}$ is the set of available rates. The derivative of this objective function with respect to $\omega_1$ can be calculated as:

$$\frac{\partial \mathcal{F}}{\partial \omega_1} = \frac{1}{|R_v|} \sum_{(i,j) \in R} (\hat{r}_{ij} - r_{ij}) \frac{\partial \hat{r}_{ij}}{\partial \omega_1} \tag{10}$$

Based on equation (8), the value of $\partial \hat{r}_{ij}/\partial \omega_1$, can be calculated as:

$$\frac{\partial \hat{r}_{ij}}{\partial \omega_1} = \frac{g \sum_{s \in U_j} Sim(s, T_i) \times r_{sj} - f \sum_{s \in U_j} Sim(s, T_i)}{g^2} \tag{11}$$

where f and g are respectively the numerator and denominator of the equation (8). The $\partial \hat{r}_{ij}/\partial \omega_0$ and $\partial \hat{r}_{ij}/\partial \omega_2$ can be derived similarly. By utilizing these derivations, the parameters can be tuned iteratively. Regarding one training rate $r_{ij}$, the parameter $\omega_1$ is updated as follows:

$$\omega_1 = \omega_1^{old} + \eta(r_{ij} - \hat{r}_{ij}) \frac{\partial \hat{r}_{ij}}{\partial \omega_1} \tag{12}$$

The complete parameters optimization procedure is presented in Figure 2.

---

1: **function** GRADIENT_BASED_PARAM_TUNING ($\Omega$, $R_T$, $R_V$, $T$) **returns** $\Omega$

▷ **Input:** $\Omega$, set of initial values for parameters $\omega_0$, $\omega_1$, and $\omega_2$

$R_T$: train rates, $R_V$: validation rates, $T$: trust network

▷ **Output:** tuned values of $\Omega$

2:  **for** tuning_iteration $j = 1$ to $n$ **do**

3:      **for** r in $R_V$ **do**

4:          calculate $\hat{r}, \frac{\partial \hat{r}}{\partial \omega_0}, \frac{\partial \hat{r}}{\partial \omega_1}, \frac{\partial \hat{r}}{\partial \omega_2}$    ▷ using $T$ and $R_T$ (see equations 8 and 11)

5:          **for** parameter i = 1 to $|\Omega|$ **do**

6:              $\omega_i = \omega_i^{old} + \eta(r - \hat{r}) \frac{\partial \hat{r}}{\partial \omega_i}$

8: **return** $\Omega$

---

**Figure 2: Parameter optimization using gradient descent scheme**

## 3. EXPERIMENTS AND EVALUATION

In this section, we first evaluate the effectiveness of using social-ties in rate-prediction based on accuracy measures and coverage. Next, we examine the contribution of social network information by learning the weight parameters. Finally, we assess the performance of our method in comparison with other methods proposed in the literature.

### 3.1. DATASETS AND EVALUATION METRICS

In this study, we employ two real-world datasets, Epinions and FilmTrust, which provide user-item ratings and explicit trust statements. The first one is the dataset of consumer review website Epinions[1], which consists of users' ratings for different items (i.e., commercial commodities) and user-user trust relationships. The rates are integer values in the range of 1-5 assigned by users to the items. The density index of the Epinions dataset which is a measure of sparsity is 0.0097% (the dataset is more than 99% sparse). The density index is the percentage of available rates over all the items to the number of possible ratings (i.e., all items rated by all of the users). The average of all rates in the Epinions dataset is 3.99, and the average number of items rated by a user is 13.49. The mean number of trusted users by each individual is 9.88.

FilmTrust is a dataset crawled from the FilmTrust website in June 2011, which consists of ratings to movies. It also provides user-user trust statements, similar to the Epinions dataset. The FilmTrust dataset is about 98.86% sparse. A user has rated 23.54 items on average and the average of all rates in the dataset is 3. Each user has trusted 1.23 users on average.

We use 5-fold cross-validation in the experiments: each time, the user-item rating set $R$ is divided into five equal folds. One fold is selected as test set, $R_E$, which is used in the evaluation phase. The remaining folds are divided into training set, $R_T$, and validation set, $R_V$. Thus, $R$ is partitioned into three disjoint sets.

**Table 1: Statistics of the datasets**

| Dataset | #User | #Item | #Rating | #Trust-relation | Density Index | Rating domain |
|---|---|---|---|---|---|---|
| Epinions | 49,289 | 139,738 | 664,824 | 487,183 | 0.0097% | [1,5] |
| FilmTrust | 1508 | 2071 | 35,497 | 1853 | 1.14% | [0.5,4] |

To evaluate the performance of the proposed methods, we have used popular metrics MAE, RMSE, Coverage, and F1. The performance metrics are defined as

$$MAE = \frac{\sum_{\hat{r}_{ui} \in \hat{R}_E} |r_{ui} - \hat{r}_{ui}|}{|\hat{R}_E|} \tag{13}$$

$$RMSE = \sqrt{\frac{\sum_{\hat{r}_{ui} \in \hat{R}_E} (r_{ui} - \hat{r}_{ui})^2}{|\hat{R}_E|}} \tag{14}$$

$$Cov = \frac{|\hat{R}_E|}{|R_E|} \tag{15}$$

---

[1] http://epinions.com

where $r_{ui}$ is the rate of item $i$ assigned by the user $u$, $\hat{r}_{ui}$ is the predicted rate corresponding $r_{ui}$, and $\hat{R}_E$ is the set of all predicted rates.

The number of predicted rates may be less than the total number of the rates available in the test set $R_E$. Specifically, condition $Sim(u, V_i) = 0 \cdot \forall i$ or $U_t = \emptyset$, in the proposed rate-prediction formula 5, causes unpredictable rates. For memory-based methods that do not guarantee the full coverage, F1 is a suitable performance metric, defined as

$$F1 = \frac{2 \times Acc \times Cov}{Acc + Cov} \quad (16)$$

where $Acc = 1 - (\sum_{r_{ui} \in R_E} |r_{ui} - \hat{r}_{ui}|/(|\hat{R}_E| \times (r_{max} - r_{min})))$.

### 3.2. IMPACT OF SOCIAL TIES

To evaluate the impact of integrating social ties into the basic CF algorithm, we consider our method without incorporating social ties ($\omega_1=\omega_2=0$ in equation (8)) as a baseline. The comparison is made by means of coverage percentage, MAE and RMSE. As defined in equation (15), the coverage metric is the fraction of the test item rates covered in the prediction.

As illustrated in table 2, for the Epinions dataset, the proposed social similarity measure reaches the coverage percentage of about 82.2%, which is 13% higher than the baseline that does not use social information (covers 69.13% of the test rates). By using only direct trust ($\omega_0=\omega_2=0$), we obtain a coverage percentage of 70.93%, which is 1.83% higher than the baseline.

In the case of the FilmTrust dataset, using only direct trust ($\omega_0=\omega_2=0$) does not provide considerable coverage (obtains a coverage percentage of 14.02%). In fact, in the trust network, only 1.23 users are trusted by each individual on average. As a result, using the proposed social similarity measure does not improve the coverage of the baseline algorithm noticeably.

**Table 2: Coverage comparison of different approaches**

| Method | Epinions | FilmTrust |
|---|---|---|
| Baseline, without social information ($\omega_1 = \omega_2 = 0$) | 69.13% | 96.75% |
| Using only direct social information ($\omega_0 = \omega_2 = 0$) | 70.93% | 13.98% |
| Direct social CF ($\omega_2 = 0$) | 79.73% | 96.83% |
| Using final similarity measure ($\omega_i \neq 0 \cdot i \in \{0,1,2\}$) | 82.2% | 96.86% |

To show the impact of using social ties on the performance of rate-prediction, first of all, we provide a simple experiment in which a single weight parameter is changed (from 0 to 1.0 with a fixed step of 0.1, while other parameters are fixed) and the effect of this change is measured using MAE and RMSE metrics.

For Epinions dataset, equation (6) is used as the similarity measure and $0 < \omega_1 = (1 - \omega_0) \leq 1$. For FilmTrust dataset, the similarity measure (8) is used in which $\omega_1 = \omega_0 = 0.5$ and $\omega_2$ is changed from 0 to 1. Since the coverage varies by changing the weight parameters, we selected the intersection of the covered subsets of the test set to provide comparable results. For example, in the case of the Epinions dataset, this test subset contains 60% of the original test set. The results on the Epinions are shown in figures 3.a and 3.b. It can be observed that using a combination of own user's opinion and social ties' opinions outperforms the non-social ($\omega_1 = 0$) and fully-social CFs ($\omega_1 = 1$), regarding both MAE and RMSE. The best results are obtained using $\omega_1 = 0.6$, where MAE and RMSE are respectively 1.1187 and 0.7875. With the increase of the weight parameter $\omega_1$, the trend of MAE and RMSE is decreasing until about 0.6 followed by a steady rise. Moreover, the simple curves of figures 3.A and 3.B show that optimal weight parameters can be simply reached by parameter tuning. The results on the FilmTrust in figures 3.C and 3.D show the complex relation of the weight parameter $\omega_2$ and the performance metrics.

To put in a nutshell, figure 3 shows that weight parameters considerably affect the performance metrics, hence parameter learning is a need. To our knowledge, learning the contribution of social data is not discussed in the literature and previous papers. In the next subsection, we discuss the result of weight parameter tuning in our similarity measure to provide the best prediction accuracy.

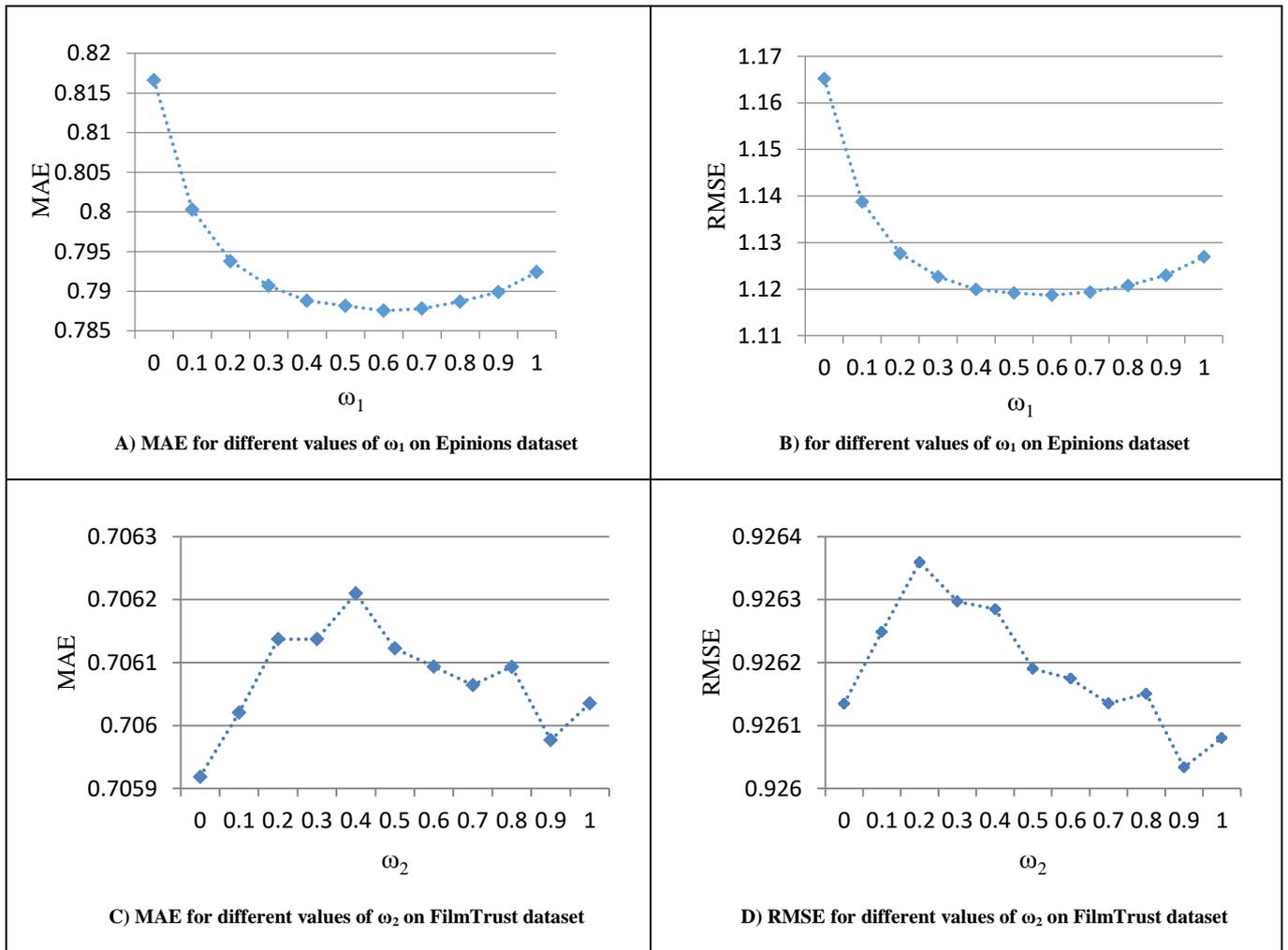

A) MAE for different values of $\omega_1$ on Epinions dataset

B) for different values of $\omega_1$ on Epinions dataset

C) MAE for different values of $\omega_2$ on FilmTrust dataset

D) RMSE for different values of $\omega_2$ on FilmTrust dataset

**Figure 3: the impact of using social ties on the performance of rate-prediction**

### 3.3. THE RESULTS OF PARAMETER LEARNING SCHEMES

The weight parameter learning procedure tunes the relative contribution of different similarity measures in equation (8). It means that the prediction performance is optimized using the best weight parameters for scorers, directly trusted users and indirectly trusted users. In the experiments, the parameters are tuned using two schemes proposed in Section 2.

The greedy algorithm of figure 1 tunes the value of one parameter in one execution. Therefore, the algorithm is repeated for tuning each of the parameters. The input arguments of the algorithm include the initial weight parameters $\Omega$, and the range of the parameter under consideration (i.e., $l$ and $u$). In the experiments, the parameters' value in $\Omega$ are initialized to 0.5 and the range $[\varepsilon, 1.0]$ is used for each parameter. It should be noted that a zero-value parameter may reduce the coverage. Hence, we bounded the minimum value to a very small number $\varepsilon=1.e-4$.

As mentioned before, the algorithm of figure 1 should be repeated for tuning each of the parameters. Since the order of parameter tuning may affect the results, we examined different orders of parameter tuning on the Epinions dataset to investigate the robustness of the algorithm. The learning procedure with each order is repeated ten times to obtain accurate results. The values of the parameters and the resultant MAE on the validation set of a certain fold are shown in Table 3. As shown in this table, the MAE on the validation set is not significantly changed. The proposed method is not sensitive to the order of parameter tuning, at least for the used datasets. To report the final results on both Epinions and FilmTrust, we used the order of $\omega_0$, $\omega_1$, and $\omega_2$, and the whole procedure is repeated three times. Then the tuned parameters on each validation set are used for the corresponding test set.

For the gradient-based algorithm of figure 2, we used Mini-batch gradient descent. In other words, the loop at line 3 is not repeated for each training instance separately but for a batch at a time. We used the batch size of 1000 for the Epinions dataset and 50 for the FilmTrust dataset. The average values of final weight parameters are summarized in Table 4.

The results of both learning schemes are reported in tables 5 and 6 for Epinions and FilmTrsut datasets respectively. In these tables, the results are compared with the baseline algorithm (i.e., not using the social information) and the greedy approach with one level of trust (i.e., using directly trusted users).

As shown in table 5, for the Epinions dataset, the greedy and gradient-based schemes obtained the same performance, based on MAE and RMSE. The proposed schemes outperform both the baseline method and the greedy learning scheme for one level of trust. The two-level learning schemes improve the performance of the baseline method for all users by reducing the MAE from 0.823 to 0.798 and the RMSE from 1.174 to 1.139. The proposed method obtains better results not only for all users' experiment but also for the cold start users.

**Table 3: MAE on the validation set with different orders of parameter learning on Epinions dataset**

| Order of Learning | Best Parameters value | | | MAE on validation |
|---|---|---|---|---|
| | $\omega_0$ | $\omega_1$ | $\omega_2$ | |
| $\omega_0, \omega_1, \omega_2$ | 0.264 | 0.324 | 0.540 | 0.80679 |
| $\omega_0, \omega_2, \omega_1$ | 0.216 | 0.268 | 0.436 | 0.80672 |
| $\omega_1, \omega_2, \omega_0$ | 0.440 | 0.516 | 0.916 | 0.80679 |
| $\omega_1, \omega_0, \omega_2$ | 0.280 | 0.364 | 0.540 | 0.80674 |
| $\omega_2, \omega_1, \omega_0$ | 0.416 | 0.516 | 0.840 | 0.80674 |
| $\omega_2, \omega_0, \omega_1$ | 0.416 | 0.516 | 0.840 | 0.80674 |

**Table 4: The tuned weight parameters for the proposed learning scheme**

| Dataset | Learning scheme | Parameters value | | |
|---|---|---|---|---|
| | | $\omega_0$ | $\omega_1$ | $\omega_2$ |
| Epinions | greedy scheme | 0.3048 ±0.0798 | 0.3608 ±0.1254 | 0.5600 ±0.0640 |
| | gradient-based scheme | 0.7456 ±0.0668 | 0.6609 ±0.0257 | 1.4194 ±0.0312 |
| FilmTrust | greedy scheme | 0.4040 ±0.4203 | 0.0741 ±0.1238 | 0.1489 ±0.2285 |
| | gradient-based scheme | 1.5935 ±0.1052 | -0.5693 ±0.1032 | 0.7006 ±0.2117 |

The trend is more perceptible for cold-start users where the proposed learning method improved the performance of the baseline method by reducing the MAE from 0.897 to 0.833 and the RMSE from 1.283 to 1.188. The rate coverage percentage of cold-start users increased by 9.83% and 29.34% in comparison with the adaptation of one level of trust and the baseline respectively.

The results on the FilmTrust dataset are shown in table 6. Similar to the results of the Epinions dataset, the performance of two learning schemes is almost equivalent. Unlike the results on the Epinions dataset, using indirect trusted users did not accompanied by tangible improvement on all users (i.e., a decrease of 0.001 for both MAE and RMSE with two-level trust gradient-based learning). The trend is similar to cold-start users where the accuracy did not change, along with an increase of 0.44% in coverage.

It should be mentioned that the social relations of the Epinions dataset consist of 9.88 trusts per user. Meanwhile, the trust network of FilmTrust is drastically sparse with 1.23 trusted individuals per user, hence less improvement utilizing social information.

Table 5: Comparison of results for different schemes of the proposed algorithms on The Epinions dataset

|  | The baseline method | | | Adapted one level trust using the greedy scheme | | | Adapted trust propagated using the greedy learning scheme | | | Adapted trust propagated using the gradient-based learning scheme | | |
|---|---|---|---|---|---|---|---|---|---|---|---|---|
|  | MAE | RMSE | Coverage | MAE | RMSE | Coverage | MAE | RMSE | Coverage | MAE | RMSE | Coverage |
| All users | 0.823 | 1.174 | 69.13 | 0.804 | 1.147 | 79.73 | 0.798 | 1.139 | 82.20 | 0.798 | 1.139 | 82.20 |
| Cold start users | 0.897 | 1.283 | 41.64 | 0.854 | 1.220 | 61.15 | 0.833 | 1.188 | 70.98 | 0.833 | 1.188 | 70.98 |

Table 6: Comparison of results for different schemes of the proposed algorithms on The FilmTrust dataset

|  | The baseline method | | | Adapted one level trust using the greedy scheme | | | Adapted trust propagated using the greedy learning scheme | | | Adapted trust propagated using the gradient-based learning scheme | | |
|---|---|---|---|---|---|---|---|---|---|---|---|---|
|  | MAE | RMSE | Coverage | MAE | RMSE | Coverage | MAE | RMSE | Coverage | MAE | RMSE | Coverage |
| All users | 0.705 | 0.925 | 96.75 | 0.705 | 0.925 | 96.83 | 0.705 | 0.925 | 96.86 | 0.704 | 0.924 | 96.86 |
| Cold start users | 0.671 | 0.882 | 94.38 | 0.670 | 0.880 | 94.73 | 0.671 | 0.881 | 94.82 | 0.671 | 0.881 | 94.82 |

The histograms of the closeness of the predicted rates and real values of rates are depicted in Figures 4 and 5 on the Epinions and FilmTrust datasets respectively. Here, the closeness of rates is defined as $|\hat{r}_{ui} - r_{ui}|$ (see equations (1) and (8)). In the case of testing for all users, about 85% of predicted rates have zero or just one difference from actual values on the Epinions dataset. The percentage of rates predicted with the maximum difference (i.e. 4) is 0.53%. In the case of testing for cold-start users, the number of zero difference rates decreases slightly. For the FilmTrust dataset, there are 8 possible difference values in the histogram. According to the histograms of Figure 5, about 89% of predicted rates are laid in the first three bins for both all users and cold start users experiments.

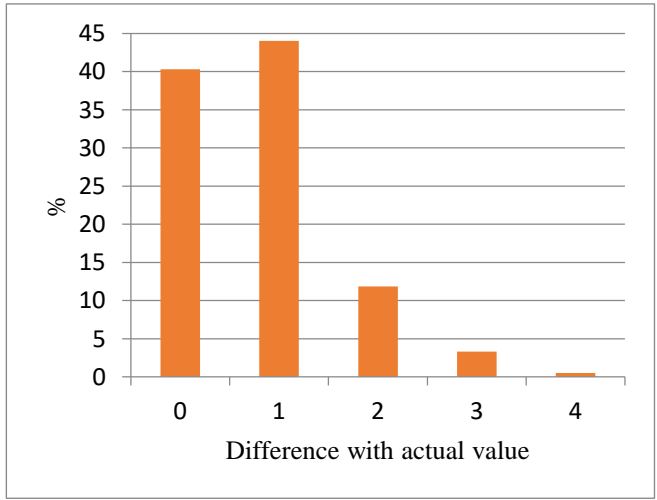

A) Histogram of predicted rates using two levels of trust for all users

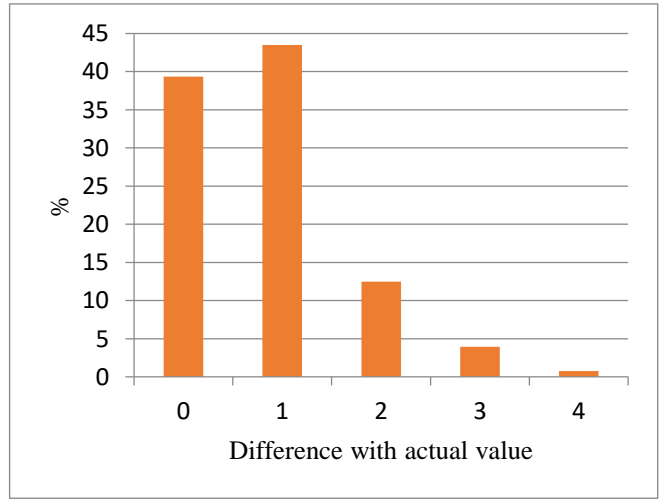

B) Histogram of predicted rates using two levels of trust for cold-start users

Figure 4: Histogram of the closeness of the predicted rates and real rates on Epinions dataset

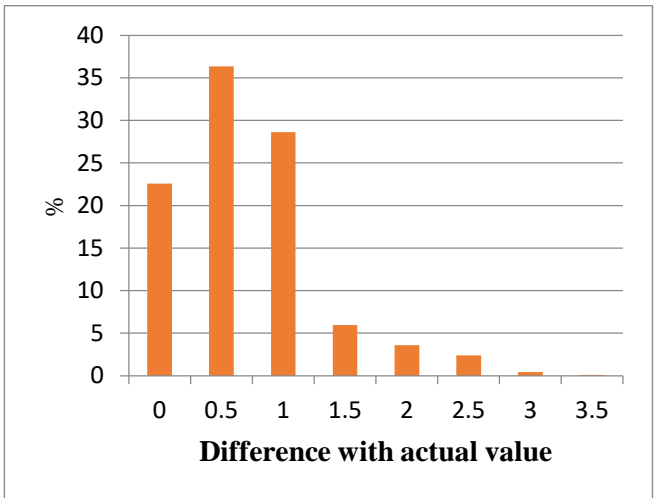 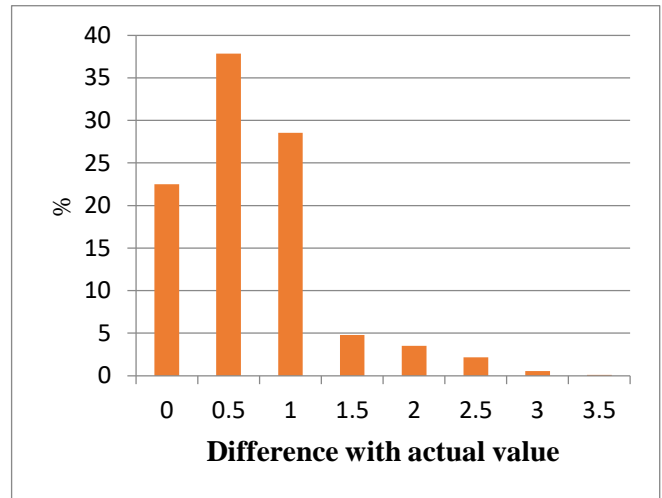

**A) Histogram of predicted rates using two levels of trust for all users**

**B) Histogram of predicted rates using two levels of trust for cold-start users**

**Figure 5: Histogram of the closeness of the predicted rates and real rates on FilmTrust dataset**

To show the efficiency and performance of our methods, we compared our results with the results of the state-of-the-art methods. Table 7 provides a comparison of the proposed method with state-of-the-art model-based methods. MAE and RMSE are used as evaluation metrics. We point out the top six methods by labels (1) to (6) in the table.

For the Epinions dataset, the best MAE belongs to the proposed trust-propagated method. The proposed method using one level of trust is ranked second and the third rank belongs to the AutoRec method. Based on RMSE, AutoRec is the winner and the TrustSVD method is the runner-up. The proposed learning schemes are placed in the third and fourth ranks.

For the FilmTrust dataset, the TrustSVD method achieved the best results. It should be noted that model-based methods are more complex than memory-based methods. The advantage of the proposed method is the simplicity and interpretability of the similarity measure, which can be easily extended by means of other similarity features.

Table 8 shows a comparison of the results of the proposed method with well-known memory-based methods. Methods TCF1 and Merge1 utilize one level of trust and methods TCF2 and Merge2 use two levels of trust. The top 5 methods are shown by numbers from (1) to (5). The proposed method obtains the best MAE results on the Epinions dataset among the other methods. The highest F1 and coverage belong to 2D-Graph clustering. Based on F1 and coverage, the runner-up method is the proposed trust propagated method. The third place is also achieved by the proposed one-level of trust method. In the case of the FilmTrust dataset, the proposed methods are in the first and second ranks based on F1 and coverage. The lowest MAE belongs to the 2D-Graph clustering method.

It is observed that our trust propagated method achieved comparable performance in spite of its simplicity and adaptability.

**Table 7: The comparison of the proposed schemes with the state-of-the-art model-based CF methods**

| Method | Datasets | | | |
|---|---|---|---|---|
| | Epinions | | FilmTrust | |
| | MAE | RMSE | MAE | RMSE |
| PMF [13] | 1.112 | 1.353 | 0.769 (5) | 1.022 |
| NMF [16] | 1.113 | 1.549 | 0.898 | 0.984 (5) |
| RSTE [29] | 1.028 (6) | 1.329 (6) | 0.857 | 1.164 |
| TrustMF [7] | 1.014 (5) | 1.273 (5) | 0.858 | 1.166 |
| TrustSVD [30] | 0.864 (4) | 1.103 (2) | 0.642 (1) | 0.825 (1) |
| AutoRec [19] | 0.837 (3) | 0.985 (1) | 0.761 (4) | 0.865 (2) |
| SoReg [27] | 1.442 | 1.688 | 0.812 (6) | 1.01 (6) |
| Adapted one level trust using | 0.804 (2) | 1.147 (4) | 0.705 (3) | 0.925 (4) |
| greedy scheme | | | | |
| Adapted trust propagated using gradient-based learning scheme | 0.798 (1) | 1.1392 (3) | 0.704 (2) | 0.924 (3) |

**Table 8: The comparison of the proposed schemes with the state-of-the-art memory-based CF methods**

| Method | Datasets | | | | | |
|---|---|---|---|---|---|---|
| | Epinions | | | FilmTrust | | |
| | MAE | F1 | Coverage | MAE | F1 | Coverage |
| TCF1 [36] | 0.867 | 0.7409 | 70.28 | 0.714 | 0.8658 | 94.92% (4) |
| TCF2 [36] | 0.864 | 0.7794 (5) | 77.48 (5) | 0.719 | 0.8661 (5) | 95.19% (3) |
| Merge1 [37] | 0.839 (5) | 0.7608 | 73.35 | 0.705 (3) | 0.8667 (4) | 94.77% |
| Merge2 [37] | 0.824 (4) | 0.7898 (4) | 78.50 (4) | 0.707 (5) | 0.8672 (3) | 94.94% (5) |
| 2D-Graph clustering [38] | 0.810 (3) | 0.830 (1) | 86.48 (1) | 0.659 (1) | 0.848 | 86.08% |
| Adapted one level trust using greedy scheme | 0.804 (2) | 0.79827 (3) | 79.73 % (3) | 0.705 (3) | 0.8902 (2) | 96.83% (2) |
| Adapted trust propagated using gradient-based learning scheme | 0.798 (1) | 0.8111 (2) | 82.20% (2) | 0.704 (2) | 0.8905 (1) | 96.86% (1) |

## 4. CONCLUSION AND FUTURE WORK

Recommender systems are used to suggest relevant data to each user. The trust concept has been recently incorporated into recommender systems to improve accuracy. In this paper, we proposed a memory-based CF method in which the query user data is extended by means of social information. For this, we introduced a similarity measure in which two levels of social trust are incorporated. We also, provided learning schemes to tune the influence of each level of trust in the similarity measure between users. Therefore, the proposed method could be simply adapted for any dataset.

The Experimental results showed the effectiveness of using social information in improving the performance of CF recommender methods from both accuracy and prediction coverage aspects.

The proposed method could be extended in future work by personalizing the weight parameters (i.e., different weights of each level of social ties for each user). More than two-level of trust propagation should be effective in this approach.

## REFERENCES

[1] J. Tang, X. Hu, and H. Liu, "Social recommendation: a review," *Social Network Analysis and Mining*, vol. 3, no. 4, pp. 1113–1133, 2013.
[2] X. Yang, Y. Guo, Y. Liu, and H. Steck, "A survey of collaborative filtering based social recommender systems," *Computer Communications*, vol. 41, pp. 1–10, Mar. 2014.
[3] M. McPherson, L. Smith-Lovin, and J. M. Cook, "Birds of a Feather: Homophily in Social Networks," *Annual Review of Sociology*, vol. 27, no. 1, pp. 415–444, Aug. 2001.
[4] F. García-Sánchez, R. Colomo-Palacios, and R. Valencia-García, "A social-semantic recommender system for advertisements," *Information Processing and Management*, vol. 57, no. 2, Mar. 2020.
[5] Y.-M. Li, C.-T. Wu, and C.-Y. Lai, "A social recommender mechanism for e-commerce: Combining similarity, trust, and relationship," *Decision Support Systems*, vol. 55, no. 3, pp. 740–752, Jun. 2013.
[6] A. Salah, N. Rogovschi, and M. Nadif, "A dynamic collaborative filtering system via a weighted clustering approach," *Neurocomputing*, vol. 175, pp. 206–215, Jan. 2016.
[7] B. Yang, Y. Lei, J. Liu, and W. Li, "Social Collaborative Filtering by Trust," *IEEE Transactions on Pattern Analysis and Machine Intelligence*, vol. 39, no. 8, pp. 1633–1647, Aug. 2017.
[8] H. Zare, M. A. Nikooie Pour, and P. Moradi, "Enhanced recommender system using predictive network approach," *Physica A:*

*Statistical Mechanics and its Applications*, vol. 520, pp. 322–337, Apr. 2019.

[9] X. Wang, S. C. H. Hoi, M. Ester, J. Bu, and C. Chen, "Learning Personalized Preference of Strong and Weak Ties for Social Recommendation," in *Proceedings of the 26th International Conference on World Wide Web - WWW '17*, 2017, pp. 1601–1610.

[10] G. Adomavicius and A. Tuzhilin, "Toward the next generation of recommender systems: a survey of the state-of-the-art and possible extensions," *IEEE Transactions on Knowledge and Data Engineering*, vol. 17, no. 6, pp. 734–749, Jun. 2005.

[11] B. M. Sarwar, B. M. Sarwar, G. Karypis, J. A. Konstan, and J. T. Riedl, "Application of Dimensionality Reduction in Recommender System -- A Case Study," *IN ACM WEBKDD WORKSHOP*, 2000.

[12] Y. Koren, R. Bell, and C. Volinsky, "Matrix Factorization Techniques for Recommender Systems," *Computer*, vol. 42, no. 8, pp. 30–37, Aug. 2009.

[13] R. Salakhutdinov and A. Mnih, "Probabilistic Matrix Factorization," in *Advances in Neural Information Processing Systems (NIPS '08)*, 2008, pp. 1257–1264.

[14] R. Salakhutdinov and A. Mnih, "Bayesian probabilistic matrix factorization using Markov chain Monte Carlo," in *Proceedings of the 25th international conference on Machine learning - ICML '08*, 2008, pp. 880–887.

[15] Y. Koren and Yehuda, "Collaborative filtering with temporal dynamics," in *Proceedings of the 15th ACM SIGKDD international conference on Knowledge discovery and data mining - KDD '09*, 2009, pp. 447–456.

[16] D. D. Lee and H. S. Seung, "Algorithms for non-negative matrix factorization," in *Advances in neural information processing systems (NIPS'00)*, 2001, pp. 556–562.

[17] N. C C and A. Mohan, "A social recommender system using deep architecture and network embedding," *Applied Intelligence*, vol. 49, no. 5, pp. 1937–1953, May 2019.

[18] Y. Koren, "Factor in the neighbors: Scalable and accurate collaborative filtering," *ACM Transactions on Knowledge Discovery from Data*, vol. 4, no. 1, pp. 1–24, Jan. 2010.

[19] S. Sedhain, A. K. Menon, S. Sanner, and L. Xie, "AutoRec," in *Proceedings of the 24th International Conference on World Wide Web - WWW '15 Companion*, 2015, pp. 111–112.

[20] M. Deshpande and G. Karypis, "Item-based top- *N* recommendation algorithms," *ACM Transactions on Information Systems*, vol. 22, no. 1, pp. 143–177, Jan. 2004.

[21] J. Wang, A. P. de Vries, and M. J. T. Reinders, "Unifying user-based and item-based collaborative filtering approaches by similarity fusion," in *Proceedings of the 29th annual international ACM SIGIR conference on Research and development in information retrieval - SIGIR '06*, 2006, p. 501.

[22] G. Linden, B. Smith, and J. York, "Amazon.com recommendations: item-to-item collaborative filtering," *IEEE Internet Computing*, vol. 7, no. 1, pp. 76–80, Jan. 2003.

[23] A. Nakamura and N. Abe, "Collaborative Filtering Using Weighted Majority Prediction Algorithms," *ICML*, vol. 98, pp. 395–403, 1998.

[24] M. R. Zarei and M. R. Moosavi, "A Social Recommender System based on Bhattacharyya Coefficient," *arXiv: 1809.03047*, Sep. 2018.

[25] Seyed Mohammad Bidoki and Seyed Mohammad Reza Moosavi, "IDUF: An active learning based scenario for relevance feedback query expansion," in *2012 International Conference on Information Retrieval & Knowledge Management*, 2012, pp. 244–248.

[26] M. Jamali and M. Ester, "A matrix factorization technique with trust propagation for recommendation in social networks," in *Proceedings of the fourth ACM conference on Recommender systems - RecSys '10*, 2010, p. 135.

[27] H. Ma, D. Zhou, C. Liu, M. R. Lyu, and I. King, "Recommender systems with social regularization," in *Proceedings of the fourth ACM international conference on Web search and data mining - WSDM '11*, 2011, p. 287.

[28] H. Ma, T. C. Zhou, M. R. Lyu, and I. King, "Improving Recommender Systems by Incorporating Social Contextual Information," *ACM Transactions on Information Systems*, vol. 29, no. 2, pp. 1–23, Apr. 2011.

[29] H. Ma, I. King, and M. R. Lyu, "Learning to recommend with social trust ensemble," in *Proceedings of the 32nd international ACM SIGIR conference on Research and development in information retrieval - SIGIR '09*, 2009, p. 203.

[30] G. Guo, J. Zhang, and N. Yorke-Smith, "A Novel Recommendation Model Regularized with User Trust and Item Ratings," *IEEE Transactions on Knowledge and Data Engineering*, vol. 28, no. 7, pp. 1607–1620, Jul. 2016.

[31] Y. Koren and Yehuda, "Factorization meets the neighborhood," in *Proceeding of the 14th ACM SIGKDD international conference on Knowledge discovery and data mining - KDD 08*, 2008, p. 426.

[32] M. Jamali and M. Ester, "TrustWalker : a random walk model for combining trust-based and item-based recommendation," in *Proceedings of the 15th ACM SIGKDD international conference on Knowledge discovery and data mining - KDD '09*, 2009, pp. 397–406.

[33] P. Massa and P. Avesani, "Trust-Aware Collaborative Filtering for Recommender Systems," Springer, Berlin, Heidelberg, 2004, pp. 492–508.

[34] P. Massa and P. Avesani, "Trust-aware recommender systems," in *Proceedings of the 2007 ACM conference on Recommender systems - RecSys '07*, 2007, p. 17.

[35] M. R. Zarei and M. R. Moosavi, "A Memory-Based Collaborative Filtering Recommender System Using Social Ties," in *4th International Conference on Pattern Recognition and Image Analysis, IPRIA 2019*, 2019, pp. 263–267.

[36] M. Chowdhury, A. Thomo, and B. Wadge, "Trust-Based Infinitesimals for Enhanced Collaborative Filtering," in *15th International Conference on Management of Data*, 2009.

[37] G. Guo, J. Zhang, and D. Thalmann, "Merging trust in collaborative filtering to alleviate data sparsity and cold start," *Knowledge-Based Systems*, vol. 57, pp. 57–68, Feb. 2014.

[38] L. Sheugh and S. H. Alizadeh, "A novel 2D-Graph clustering method based on trust and similarity measures to enhance accuracy and coverage in recommender systems," *Information Sciences*, vol. 432, pp. 210–230, Mar. 2018.